\documentstyle[psfig, referee]{mn}
\topmargin -0.5cm \title[]{Overall Temporal Synchrotron Emissions from Relativistic Jets:
Adiabatic and Radiative Breaks}

\author[]{Zhuo Li\thanks{lizhuo@nju.edu.cn},
                  Z. G. Dai\thanks{daizigao@public1.ptt.js.cn}
          and T. Lu\thanks{tlu@nju.edu.cn}\\
\rm Department of Astronomy, Nanjing University, Nanjing 210093, China}

\date{(submitted to {\em MNRAS})}
\pubyear{2001}

\begin{document}

\maketitle

\begin{abstract}
We discuss the afterglow emission from a relativistic jet that is
initially in the radiative regime in which the accelerated
electrons are fast cooling. We note that such a ``semiradiative''
jet decelerates faster than an adiabatic jet does. We also take
into account the effect of strong inverse-Compton scattering on
the cooling frequency in the synchrotron component and therefore
on the light curve decay index. We find that there are two kinds
of light-curve break for the jet effect. The first is an
``adiabatic break'' if the electrons become slow cooling before
the jet enters a spreading phase, and the second is a ``radiative
break'' which appears on the contrary case. We then show how a
relativistic jet evolves dynamically and derive the overall
temporal synchrotron emission in both cases, focusing on the
change in light curve decay index around the break time. Finally,
in view of our results, we rule out two cases for relativistic
jets to account for the observed light curve breaks in a few
afterglows: (i) an adiabatic jet with strong Compton cooling
($Y>1$) and with the cooling frequency $\nu_c$ locating in the
observed energy range; (ii) a radiative jet with a significant
fraction of total energy occupied by electrons ($\epsilon_e \sim
1$).
\end{abstract}


\section{Introduction}

It is widely believed that gamma-ray bursts (GRBs) and their afterglows
are caused by the dissipation of kinetic energy of an ultrarelativistic ejecta,
with Lorentz factor $\gamma>100$, releasing from the central engine
(so-called ``fireball" model, see Piran 1999 and van Paradijs, Kouveliotou
\& Wijers 2000 for detailed reviews). GRBs are the most energetic
explosive phenomena in astronomy. In the fireball
model, the collisions among different shells within the ejecta
might be expected to produce a prompt
burst (Paczy{\'n}ski \& Xu 1994; Rees \& M\'esz\'aros 1994),
and subsequently the ejecta interacts with its surrounding medium,
producing a long-term and broad-band afterglow (Paczy{\'n}ski \&
Rhoads 1993; M\'esz\'aros \& Rees 1997). In the standard afterglow
picture, the ejecta drives a relativistic blast wave expanding into
the surrounding medium, which approaches a self-similar solution
(Blandford \& McKee 1976) after a short time
once the swept-up medium attains an energy comparable to
the initial energy of the burst.  Electrons in the cold medium are heated
in the shock front to highly relativistic energy and produce
a broad-band afterglow via synchrotron/inverse-Compton emission.
The predicted emission spectrum
and light curve (M\'esz\'aros \& Rees 1997; Sari et al. 1998)
have met essential successes in describing the behavior of
afterglows (e.g., Vietri 1997; Waxman 1997ab; Wijers et al. 1997).
The recent observations on GRB afterglows have lead to numerous
numerical calculations to model the observed afterglows
(e.g., Huang et al. 2000a, b; Gou et al. 2001; Panaitescu \& Kumar 2001;
Panaitescu 2001).

We re-investigate the dynamical evolution and temporal emission
of afterglows by considering three aspects.
(i) The electrons heated by the shock are always fast cooling when the
cooling timescale for these electrons is shorter than the dynamical
timescale (see, e.g., Waxman 1997a; M\'esz\'aros et al. 1997;
Sari et al. 1998), and the afterglow evolution should be in the radiative
regime rather than the adiabatic one as long as
a significant fraction ($\epsilon_e$) of the fireball energy is
transferred to newly heated electrons, $\epsilon_e \sim 1$.
This shock decelerates faster than an adiabatic
shock, as pointed out by B\"ottcher \& Dermer (2000).
The newly launched HETE-2 satellite, due to its rapid and accurate
location of GRBs, should lead to much more
follow-up observations on early X-ray and optical emission,
allowing a detailed analysis of the afterglow feature.
(ii)  The inverse-Compton emission is important in afterglows (Panaitescu
\& M\'esz\'aros 1998; Wei \& Lu 1998; Totani 1998; Chiang \& Dermer 1999;
Dermer, B\"ottcher \& Chiang 2000;
Panaitescu \& Kumar 2000; Sari \& Esin 2001; Zhang \& M\'esz\'aros 2001).
When the inverse-Compton cooling of
electrons dominates the synchrotron cooling, the temporal synchrotron
emission is influenced by changing the scaling law of the cooling frequency
$\nu_c$ (defined later) with time.
(iii) Jets are especially important in GRBs since they are relevant to almost all
aspects of GRB phenomenon, e.g. the total energy released in a GRB, the
burst rate, the ejection mechanism of the central engine and the dynamical evolution
of the ejecta. It is of particular interest that the light curves of afterglows
might be changed when the jet enters a spreading phase,
as pointed out by Rhoads (1999), Sari et al. (1999) and Dai, Huang \& Lu (2001).
The discovery of polarization in the afterglow of GRB 990510 has shown evidence for
a jet-like outflow (Covino et al. 1999; Wijers et al. 1999).

We consider GRBs as jets. The jet evolves
as spherical-like expansion before entering the spreading phase
(Rhoads 1999; Sari et al. 1999), and the heated electrons
are always fast cooling at first. This radiative and  spherical-like
phase ends once either (1) the electrons
become slow cooling or (2) the relativistic jet transits to the spreading phase.
So a ``radiative break'' in the light curve may appear for the case
where the jet transits to the spreading phase when the electrons are still fast
cooling and the jet is in the radiative regime. This break is
different from the widely discussed ``adiabatic break'' for the case
where the transition occurs in the adiabatic regime in which the electrons have been
slow cooling. In this work we derive the overall analytical evolution
of synchrotron emission from a relativistic jet. We first introduce
the dynamical evolution of a relativistic jet in section 2. Then the temporal
evolution of synchrotron emission from the jet is calculated in section 3.
In section 3.1 we analyze the radiative regime when the jet evolution
is spherical-like. Section 3.2 discusses the adiabatic break in which the
electrons become slow cooling before the jet evolves into the spreading phase
while section 3.3 explores the radiative break in which the transition to
the spreading phase occurs in the radiative regime. We give some discussions
and conclusions in section 4.

\section{Dynamical evolution of relativistic jets}

Let's consider a relativistic jet expanding into its surrounding medium with an initial
half opening angle of $\theta_0$, a laterally-spreading velocity of $c_s$ and a
bulk Lorentz factor of $\gamma$. Though the shock is beamed, the jet evolution
is spherical-like as long as $\gamma>\theta_0^{-1}(c_s/c)$. Because the electrons
accelerated by a shock are always fast cooling at first, the jet should be radiative
if the energy density of the electrons is a significant fraction $\epsilon_e$
of the total energy density of the shocked medium. The jet hydrodynamics
in this radiative regime will be different from the well known self-similar solution
for adiabatic blast waves derived by Blandford \& McKee (1976),
$\gamma\propto r^{-3/2}$, where $r$ is the shock radius.
Thus, the jet behaves as a radiative and spherical-like expansion
at the initial stage. The jet evolution later is divided into two cases.
({\it a}) In the case of an
adiabatic break, the accelerated electrons become slow cooling before
the jet evolves into a spreading phase, so the jet exhibits an adiabatic and
spherical-like evolution as $\gamma\propto r^{-3/2}$ sequentially and finally a spreading
phase of $\gamma$ dropping exponentially with increasing $r$.
({\it b}) In the case of a
radiative break, the jet transition into the spreading phase occurs in the radiative
regime where the shocked electrons are still fast cooling, and
therefore the jet evolves from the initial radiative spherical-like phase
to the spreading phase directly, without the second phase of adiabatic and
spherical-like evolution of case ({\it a}). Fig. 1 depicts these two cases of
the dynamical evolutions of relativistic jets.

Since the evolution of an adiabatic spherical blast wave is
well known, we introduce here only the dynamics of so-called
``semiradiative'' blast waves for the initial stage of radiative and spherical-like
evolution and of the radiative jet for the final stage in the case of a radiative
break.

Recently, for semiradiative blast waves, in which the energy radiated is considerable to
affect the evolution of the blast waves, Cohen et al. (1998) have also derived a self-similar
solution under the assumption that a fixed fraction of the energy generated
by the shock is radiated away. This assumption is applicable
at early times, since the heated electrons are in the fast-cooling regime and
a fixed fraction of the total energy that has been transferred to electrons is radiated away
while the protons do not radiate their energy and remain hot.
Following Cohen et al. (1998), the radius $r$ and
the Lorentz factor $\gamma$ of a semiradiative
blast wave are in relation of $\gamma^2\propto r^{-m}$,
and they evolve with the observer time as
\begin{equation}
r=r_0 \left ( \frac{t}{t_0} \right ) ^{\frac{1}{m+1}},
\end{equation}
\begin{equation}
\gamma=\gamma_0 \left ( \frac{t}{t_0} \right ) ^{-\frac{m}{2(m+1)}},
\end{equation}
\begin{equation}
m=\frac{(1+\epsilon)^2+3(1+\epsilon)(4-k)-4}{3-\epsilon}.
\end{equation}
Here $\gamma_0$ is the initial coasting Lorentz factor of the jet,
$r_0$ is the deceleration radius where the energy in the shocked
medium equals that in the original explosion, $t_0$ is the observer time
at which the deceleration radius is reached, $\epsilon$ is
a fraction of energy that is radiated away and
should be understood as $\epsilon=\epsilon_e$ in the fast cooling regime,
and $k$ parameterizes the proton number density
$n$ of the surrounding medium with $n=Ar^{-k}$. For $k=0$ and
$\epsilon$=(0---1), $m$ varies from 3 to 12 and the Lorentz factor of
the blast wave evolves based on $\gamma \propto t^{-3/8}$ to
$\gamma \propto t^{-6/13}$. It is the adiabatic solution case with $\epsilon=0$.
Denoting the energy in the original explosion as $E_0$, and the redshift
of this explosion as $z$, we have
\begin{equation}
r_0=\left (\frac {3-k}{4\pi A}\frac{E_0}{\gamma_0^2m_pc^2} \right )^{\frac 1{3-k}},
\end{equation}
\begin{equation}
t_0=\frac {r_0(1+z)}{2\gamma_0^2c}.
\end{equation}

In the fully radiative ($\epsilon =1$) case, the scaling $\gamma
\propto r^{-6}$ differs from the widely used Blandford-McKee
solution, $\gamma \propto r^{-3}$ (see, e.g., Sari, Piran \&
Narayan 1998, B\"ottcher \& Dermer 2000), which assumes a thin
shell with a cooled interior and the kinetic energy of the shell
can be written as $E_k=(\gamma -1)M$, where $M$ is the shell mass.
In fact, the quantities, such as velocity, density and energy
density, should be a function of distance from the shock front.
Thus, we prefer to use the self-similar solution of Cohen et al.
(1998) which assumes a shell with finite width within which the
interior exerting a pressure onto the outer. Here after we use the
equations above to describe the evolution of blast waves in the
following calculations.

For the dynamics of a relativistic jet,
Rhoads (1999) has derived an adiabatic solution, in which the jet evolution in
the spreading phase is an exponential slowing down of $\gamma$ with increasing $r$.
We can expect the jet slows down even faster if it is in the radiative regime and
loses energy more quickly than an adiabatic jet does.
Thus, for a radiative jet, $r$ is practically
a constant during the spreading phase. Therefore, from $r\propto \gamma^2 ct$,
we have the same scaling relation of $\gamma\propto t^{-1/2}$ as that of an
adiabatic jet.

\section{Temporal evolution of synchrotron emission spectra}

As shown later, the previous studies on jetted afterglows are only suitable for
the adiabatic break case with weak Compton-cooling effect.
At first, we concentrate on deriving the spectra and light curves
in the initial radiative and spherical-like stage, and then we discuss
the light-curve change around the break time in both the adiabatic and radiative
break cases with strong or weak Compton-cooling effect.

\subsection{Radiative and spherical-like phases}
We only consider the emission from the accelerated electrons,
neglecting the emission from the hot protons. The electron
distribution is assumed to be a power-law of Lorentz factor,
$N(\gamma_e)\propto\gamma_e^{-p}$, with the minimum random Lorentz
factor,
\begin{equation}
\gamma_m=\frac 2{(1+X)}\frac{(p-2)}{(p-1)}\frac{m_p}{m_e}\epsilon_e\gamma,
\end{equation}
where $X$ is the usual hydrogen mass fraction of the surrounding medium.
We assume that $p>2$ here after so that the total electron energy is dominated by
electrons with $\gamma_m$ [Recently Dai \& Cheng (2001)
have studied the case of $1<p<2$].
The strength of the magnetic field in the co-moving frame is given by
\begin{equation}
B'=\gamma c\sqrt{32\pi nm_p\epsilon_B},
\end{equation}
where $\epsilon_B$ is a fraction of the total internal energy densty that is
carried by the magnetic field.

We concentrate here on the radiative regime when the cooling timescale
for electron with $\gamma_m$ is shorter than the dynamical timescale, $t=\frac
{1+z}{c}\int\frac
{dr}{\gamma}=\frac{2(1+z)}{2+m}\frac r{c\gamma}$ for $\gamma\propto r^{-m/2}$. In this
case, the distribution of electrons extends to a lower Lorentz factor $\gamma_c$
as a power-law $\gamma_e^{-2}$ for $\gamma_c<\gamma_e<\gamma_m$ and $\gamma
^{-(p+1)}$ for $\gamma_e>\gamma_m$, where $\gamma_c$ is the Lorentz factor of electrons
that cool on the dynamical timescale $t$,
\begin{equation}
\gamma_c=\frac{3\pi m_ec^2}{\sigma_T}\frac{(2+m)}{(1+Y)}\frac{\gamma}{B'^2r}.
\end{equation}
Here we have considered the inverse-Compton cooling of electrons
by the Compton parameter (Sari \& Esin 2000)
\begin{equation}
Y=-\frac 1 2+\frac 1 2\sqrt{1+4\frac{\eta\epsilon_e}{\epsilon_B}}\simeq
\left\{ \begin{array}{ll}
  \frac{\eta\epsilon_e}{\epsilon_B},  &{\rm if }~\frac{\eta\epsilon_e}{\epsilon_B}\ll 1,\\
  \sqrt{\frac{\eta\epsilon_e}{\epsilon_B}}, &{\rm if }~\frac{\eta\epsilon_e}{\epsilon_B}\gg 1,
                             \end{array} \right.
\end{equation}
where $\eta$ is a fraction of the electron energy that is radiated
away. We should note that $\eta=1$ for fast cooling and $\eta=(\gamma_c/\gamma_m)^
{2-p}$ for slow cooling.

The simple model of synchrotron spectra of afterglows is
a broken power-law with three break
frequencies. One is the absorption frequency $\nu_a$ below which the synchrotron
photons are self-absorbed by electrons. The other two are the characteristic synchrotron
frequencies according to electrons with Lorentz factor $\gamma_m$ and $\gamma_c$,
denoted $\nu_m$ and $\nu_c$, respectively.
Following the improved treatment of synchrotron emission by Wijers \& Galama
(1999), we have
\begin{equation}
\nu_m=\frac{x_pe}{\pi m_ec(1+z)}B'\gamma_m^2\gamma,
\end{equation}
\begin{equation}
\nu_c=\frac{0.286e}{\pi m_ec(1+z)}B'\gamma_c^2\gamma,
\end{equation}
where the dimensionless factor $x_p\simeq0.64$, for $p=2$.

The absorption frequency is defined to be the point where $\tau_{\nu}
=1$. Note that in the co-moving frame of the shocked shell (denoted with a prime),
the absorption coefficient $\alpha'_{\nu'}$ scales as $\alpha'_{\nu'}
\propto\nu'^{-(p'+4)/2}$ for $\nu'>\nu_p'\equiv {\rm min}(\nu_m',\nu_c')$ and
$\alpha'_{\nu'}\propto\nu'^{-5/3}$
for $\nu'<\nu_p'$ (Waxman 1997b),
where $p'$ is the index of the distribution of those electrons
relevant to synchrotron self-absorption around $\nu'$. For the
absorption coefficient at co-moving cooling frequency
$\nu_c'=3(1+z)\nu_c/(4\gamma)$
we should take $p'=2$ and then (Rybicki \& Lightman 1979)
\begin{equation}
\alpha'_{\nu_c'}=\frac{\sqrt{3}e^3}{8\pi m_e}\left(\frac{3e}{2\pi m_e^2c^3}
\right)C\lambda B'^{2}\nu_c'^{-3}\Gamma(4)\Gamma(2),
\end{equation}
where $C=2(1+X)n\gamma\gamma_c$
(coming from $\int_{\gamma_c}^{\gamma_m}C\gamma_e^{-2}d\gamma_e
=\frac{1+X}{2}4\gamma n$ for $\gamma_m\gg\gamma_c$), $\lambda=\int_0^{\pi/2}(\sin\alpha)^2
\sin{\alpha}{\rm d}\alpha$ and $\Gamma (y)$ is the Gamma function.
The optical depth at $\nu_c'$ is then $\tau_c
\equiv\alpha'_{\nu_c'}\Delta r'$, where $\Delta r'=r/4\gamma$ is
the co-moving width of the shocked shell,
thus we can get the absorption frequency
\begin{equation}
\nu_a=\nu_c\tau_c^{3/5}=\nu_c\left (\frac{\alpha'_{\nu_c'}r}{4\gamma}\right )^{3/5}
\end{equation}
for $\nu_a<\nu_c$ in the radiative regime. We have neglected the case of $\nu_a>\nu_c$
since $\tau_c>1$ lasts for only few seconds in general parameter space.

The peak flux in the synchrotron spectrum is
\begin{equation}
F_m=\frac{\sqrt{3}\phi_pe^3}{4\pi d_l^2(1+z)m_ec^2}\gamma B'N_e,
\end{equation}
where $\phi_p$ is a factor defined by Wijers \& Galama (1999), which is
$\phi_p=0.64$ for $p=2$ here, $d_l$ is the luminosity distance, and $N_e$ is the number of
electrons in the shocked shell and given by
\begin{equation}
N_e=\frac{2\pi(1+X)}{3-k}Ar^{3-k}.
\end{equation}

Substituting $\gamma$ and $r$ with aid of equations (1)-(3), one obtains
expressions of $\nu_m$, $\nu_c$, $\nu_a$ and $F_m$ (shown in Appendix), and
their scaling relations with the observer's time are
\begin{equation}
\nu_m\propto t^{-\frac{k+4m}{2(1+m)}},
\end{equation}
\begin{equation}
\nu_c\propto t^{-\frac{4-3k}{2(1+m)}},
\end{equation}
\begin{equation}
\nu_a\propto t^{-\frac{9k+6m-8}{5(1+m)}},
\end{equation}
\begin{equation}
F_m\propto t^{-\frac{3k+2m-6}{2(1+m)}}.
\end{equation}
Given these three break frequency and peak flux,
we can calculate the synchrotron spectrum and light curve. Note that
$\nu_a<\nu_c<\nu_m$ in the radiative regime. For a homogeneous medium, $k=0$
(Note that hereafter in this paper the discussions are all concerned about
this homogeneous medium case),
the fluxes in the four frequency ranges divided by three break frequencies
evolve as
\begin{equation}
\begin{array}{l}
F_{\nu < \nu_a}=F_m\left(\frac{\nu}{\nu_a}\right)^2 \left(\frac{\nu_a}{\nu_c}
\right)^{1/3}\propto \nu^{2}t,\\
F_{\nu_a<\nu<\nu_c}=F_m\left(\frac{\nu}{\nu_c}\right)^{1/3}  \propto
                             \nu^{1/3}t^{-\frac{3m-11}{3(1+m)}},\\
F_{\nu_c<\nu<\nu_m}=F_m\left(\frac{\nu}{\nu_c}\right)^{-1/2}
                              \propto \nu^{-1/2}t^{-\frac{m-2}{1+m}},\\
F_{\nu>\nu_m}=F_m\left(\frac{\nu}{\nu_m}\right)^{-p/2}\left(\frac{\nu_m}{\nu_c}\right)^{-1/2}
                               \propto \nu^{-p/2}t^{-\frac{mp-2}{1+m}}.
\end{array}
\end{equation}
The light curve decay indexes are relevant to $\epsilon(=\epsilon_e)$, and shown in Fig. 2.
Therefore, observations of early-time afterglows
following a prompt burst can help to determine the parameter $\epsilon_e$.

We have neglected the effect of inverse-Compton process on the shape of synchrotron
spectrum, because the Thomson optical depth of synchrotron photons is generally very
small and only a negligibly small fraction of synchrotron photons is scattered. Besides,
the typical energy of inverse-Compton emission, $\nu_m^{IC}\sim\gamma_m^2\nu_m$,
usually far exceeds that of the synchrotron component which we are concerned about,
since the accelerated electrons are always relativistic because
$\gamma_m\gg\gamma\gg 1$ for relativistic blast waves.

The end of this radiative and spherical-like stage is determined
by the minimum of two time scales. One is the time $t_{cm}$ when
$\nu_m=\nu_c$. After this time, the accelerated electrons around
$\gamma_{m}$ become slow cooling so that the shock is practically
not radiative any more and transits from the radiative to the
adiabatic phase (The transition is gradual due to the gradual
change of $\epsilon$ from $\epsilon_e$ to 0, since $\epsilon =
\epsilon_e \, \eta$. Thus, in this transition phase, the blast
wave evolution will not be self-similar). Another is the break
time $t_{jet}$ when $\gamma\sim\theta_0^{-1}(c_s/c)$ and the jet
enters the spreading phase. Having the expressions of $\nu_m$ and
$\nu_c$ we can derive $t_{cm}$ (a detailed expression given in
Appendix). With the aid of the evolution of $\gamma$, we have the
break time
\begin{equation}
t_{jet}=t_0(\gamma_0\theta_0)^{\frac{2(m+1)}{m}}
\end{equation}
(see a detail in Appendix).
Here we have taken $c_s\sim c$. Fig. 3 shows two times, $t_{cm}$
and $t_{jet}$, varying with $\epsilon_e$. It is the adiabatic break case in the
small $\epsilon_e$ end since $t_{cm}<t_{jet}$, while the radiative
break case tends to appear in the right end where $t_{cm}>t_{jet}$.

\subsection{Adiabatic jet break}
\subsubsection{Adiabatic and spherical-like phase}
We first discuss the case of $t_{cm}<t_{jet}$, which corresponds to Fig. 1a.
When the time $t_{cm}$ is reached, the jet enters
an adiabatic and spherical-like stage which have been studied by many authors.
The blast wave evolves as $\gamma\propto r^{-3/2}\propto t^{-3/8}$. We have
the well-known scaling relations
\begin{equation}
\nu_m\propto t^{-3/2},~\nu_c\propto t^{-1/2}~(Y<1),~\nu_a\propto t^0,~F_m\propto t^0.
\end{equation}
In some $\epsilon_B \ll \epsilon \ll 1$ cases, i.e. a magnetic
field extremely far below equipartition like the cases of GRB
971214 (Wijers \& Galama 1999) and 990123 (Galama et al. 1999),
the adiabatic blast wave will be strong Compton cooling with $Y\gg
1$. The evolution of the cooling frequency $\nu_c$ becomes
somewhat complicated if the $Y>1$ case is considered. We combine
$Y\propto \sqrt{\eta}\propto(\gamma_c/\gamma_m)^{(2-p)/2}$ and
$\gamma_c\propto \gamma/(YB'^2r)$ for $Y>1$ with $\nu_c\propto
B'\gamma\gamma_c^2$ to obtain the general scaling
\begin{equation}
\nu_c\propto\gamma^{2-2p/(4-p)}r^{-4/(4-p)}.
\end{equation}
For an adiabatic and spherical-like jet, this becomes
\begin{equation}
\nu_c\propto t^{-3/2+2/(4-p)}~(Y>1).
\end{equation}
We can see that $\nu_c$ might increase or decrease with time,
depending on whether $p>8/3$ or $p<8/3$, respectively. Note that
$Y$ is also a function of time due to decreasing $\eta$. So if
$\epsilon_e\gg\epsilon_B$, then $Y>1$ in the beginning fast
cooling phase, and $Y\propto(\gamma^2r)^{\frac{p-2}{4-p}}$
generally in the slow cooling phase. The slowing-down leads to
$Y\propto t^{-{1\over2}(\frac{p-2}{4-p})}$ for $\gamma\propto
t^{-3/8}$, or $Y\propto t^{-\frac{p-2}{4-p}}$ for $\gamma\propto
t^{-1/2}$ until $Y\sim 1$ (We assume $p<4$ here after, which is
consistent with the observations which show that all afterglows
appear to have the electron energy distribution of $p<4$). Thus
there might be a transition from the strong Compton cooling
($Y>1$) to the weak Compton cooling ($Y<1$) phase, as pointed out
by Sari et al. (2000). However, if $\epsilon_e\ll\epsilon_B$, then
$Y<1$ throughout.

It is now easy to derive the light curves in four frequency
ranges
\begin{equation}
\begin{array}{l}
F_{\nu<\nu_a}=F_m\left(\frac{\nu}{\nu_a}\right)^2
\left(\frac{\nu_a}{\nu_m}\right)^{1/3}\propto \nu^{2}t^{1/2},\\
F_{\nu_a<\nu<\nu_m}=F_m\left(\frac{\nu}{\nu_m}
\right)^{1/3}\propto \nu^{1/3}t^{1/2},\\
F_{\nu_m<\nu<\nu_c}=F_m\left(\frac{\nu}{\nu_m}
\right)^{-(p-1)/2}\propto \nu^{-(p-1)/2}t^{-3(p-1)/4},\\
F_{\nu>\nu_c}=F_m\left(\frac{\nu}{\nu_c}\right)^{-p/2}
\left(\frac{\nu_c}{\nu_m}\right)^{-(p-1)/2} \propto \left\{ \begin{array}{ll}
  \nu^{-p/2}t^{-3p/4+1/(4-p)} &(Y>1),\\
  \nu^{-p/2}t^{-3p/4+1/2} &(Y<1).
                             \end{array} \right.
                             \end{array}
\end{equation}
Note that the transition of a jet from the radiative phase to the
adiabatic phase at $t_{cm}$
results in a flattening of the light curve in the energy range of $\nu>\nu_c$,
and even the light curve is more flattening in the case of $Y>1$ than in the case
of $Y<1$.

\subsubsection{Spreading phase}
A break time is reached when $\gamma\sim\theta_0^{-1}(c_s/c)$. In this
adiabatic break case,
Eq. (21) is invalid to calculate this break time since an adiabatic
and spherical-like phase has appeared before the final spreading phase.
We need to derive the break time again as follows. From the integral $t_{jet}^a-t_{cm}=
\int_{r_{cm}}^{r_{jet}}\frac{dr}{2\gamma^2c}$ where $r_{jet}=(\gamma_{cm}
\theta_0)^{2/3}r_{cm}$, the break time is given by
\begin{equation}
t_{jet}^a\simeq \frac{1+m}4(\gamma_{cm}\theta_0)^{8/3}t_{cm}
\end{equation}
(see a detail in Appendix)
for $t_{jet}^a\gg t_{cm}$ or $\gamma_{cm}\theta_0\gg 1$. This value is larger
than that of Eq. (21).

Having entered the spreading phase, the jet evolves as $\gamma\propto t^{-1/2}$
and the shock radius $r$ can be regarded as a constant. We rewrite the scalings of
the break frequencies and the peak flux as
\begin{equation}
\nu_m\propto t^{-2},~\nu_c \propto \left\{ \begin{array}{ll}
 t^{-2+4/(4-p)}&(Y>1)\\ t^0&(Y<1) \end{array}\right.\;,
~\nu_a\propto t^{-1/5},~F_m\propto t^{-1}.
\end{equation}
It is a rather remarkable result that $\nu_c$ is actually
increasing with time in the $Y > 1$ case. Therefore the flux turns
to evolve as
\begin{equation}
\begin{array}{l}
F_{\nu<\nu_a}\propto \nu^{2}t^0,\\
F_{\nu_a<\nu<\nu_m}\propto \nu^{1/3}t^{-1/3},\\
F_{\nu_m<\nu<\nu_c}\propto \nu^{-(p-1)/2}t^{-p},\\
F_{\nu>\nu_c}\propto \left\{ \begin{array}{ll}
  \nu^{-p/2}t^{-p+2/(4-p)}&(Y>1),\\
  \nu^{-p/2}t^{-p}&(Y<1).
                             \end{array} \right.
                             \end{array}
\end{equation}
Note that if $Y<1$, the transition of a relativistic jet into
the spreading phase results in a steepening of the light curve in $\nu>\nu_m$,
with decay index $p$ independent of whether $\nu>\nu_c$ or $\nu<\nu_c$.
But in the case of strong Compton cooling, $Y>1$, something different
happens: the transition yields the same steepening of the light curve into
$t^{-p}$ in $\nu_m<\nu<\nu_c$,
but due to the cooling frequency
$\nu_c$ increasing in time, a flattening from $t^{-3p/4+1/(4-p)}$
to $t^{-p+2/(4-p)}$ appears in $\nu>\nu_c$ .
Define the light curve index $\alpha$ and the
spectral index $\beta$ as $F_{\nu}(t)\propto t^{-\alpha}\nu^{-\beta}$.
Table 1 summarizes the relations between $\alpha$ and $\beta$ above
$\nu_m$ for different cases. Fig. 4 presents the effect of strong
Compton cooling on the $\alpha-\beta$ relation above $\nu_c$.

\begin{center}
\begin{table*}
\begin{center}
\begin{tabular}{|c||c||c|c|}
\hline
& spectral index & \multicolumn{2}{|c|}{light curve index $\alpha$
($F_{\nu}\propto t^{-\alpha}$)} \\
& $\beta$ ($F_{\nu}\propto \nu^{-\beta}$) & sphere & jet \\ \hline\hline
&  & $\alpha=3(p-1)/4$ & $\alpha=p$ \\
\raisebox{1.5ex}[0pt]{$\nu_m<\nu<\nu_c$} & \raisebox{1.5ex}[0pt]{$(p-1)/2$}
& $\alpha=3\beta/2$ & $\alpha=2\beta+1$ \\ \hline
&  & $\alpha=3p/4-1/2$ & $\alpha=p$ \\
\raisebox{1.5ex}[0pt]{$\nu>\nu_c$, $Y<1$} & \raisebox{1.5ex}[0pt]{$p/2$}
& $\alpha=3\beta/2-1/2$ & $\alpha=2\beta$ \\ \hline
&  & $\alpha=3p/4-1/(4-p)$ & $\alpha=p-2/(4-p)$ \\
\raisebox{1.5ex}[0pt]{$\nu>\nu_c$, $Y>1$} & \raisebox{1.5ex}[0pt]{$p/2$}
& $\alpha=3\beta/2-1/(4-2\beta)$ & $\alpha=2\beta-1/(2-\beta)$ \\ \hline
\end{tabular}
\end{center}
\par
\label{t:afterglow}
\caption{The spectral index $\beta$ and the light curve index $\alpha$ as
function of $p$ in the case of an adiabatic jet break. The parameter-free
relation between $\alpha$ and $\beta$ is given for each case
by eliminating $p$.}
\end{table*}
\end{center}

\subsection{Radiative jet break}
If $t_{jet}<t_{cm}$, corresponding to Fig. 1b,
the break time is first reached to end the spherical-like
evolution even when the electrons are still fast cooling and the jet might be
in the radiative regime with $\epsilon_e\sim 1$. In the spreading phase
the effect of sideways expansion dominates the jet evolution, then the
dynamics is $\gamma\propto t^{-1/2}$. And the electrons turns from fast
cooling into slow cooling in the spreading phase.

\subsubsection{Fast cooling}
With the same jet evolution as that in the spreading phase of an adiabatic jet,
$\gamma\propto t^{-1/2}$,
we have the same scalings of $\nu_m$, $\nu_c$ and $F_m$,
\begin{equation}
\nu_m\propto t^{-2},~\nu_c\propto t^0,~F_m\propto t^{-1}.
\end{equation}
Since $\eta=1$ for fast cooling, the Compton parameter $Y$ is then
a constant and the scaling relation of $\nu_c$ with observer's
time is the same as in the case of $Y\ll 1$. Next we derive the
remaining absorption frequency $\nu_a$. In this case of fast
cooling, the electrons responsible for low energy emission are
those with $\nu_c$, and therefore $\nu_a=\nu_c\tau_c^{3/5}$. In
the co-moving frame of the shocked medium, the absorption
coefficient at cooling frequency is given by Eq. (12),
$\alpha'_{\nu_c'}\propto CB'^2\nu_c'^{-3}\propto r^5\gamma^5$, and
the optical depth at cooling frequency is
$\tau_c\propto\alpha'_{\nu_c'}r/\gamma \propto r^6\gamma^4$. Thus,
we have
\begin{equation}
\nu_a \propto r^{8/5}\gamma^{12/5}\propto t^{-6/5}.
\end{equation}
With the scalings above, the spectra in fast cooling phase, where
$\nu_a<\nu_c<\nu_m$, evolve as
\begin{equation}
\begin{array}{l}
F_{\nu<\nu_a}\propto \nu^{2}t,\\
F_{\nu_a<\nu<\nu_c}\propto \nu^{1/3}t^{-1},\\
F_{\nu_c<\nu<\nu_m}\propto \nu^{-{1/2}}t^{-1},\\
F_{\nu>\nu_m}\propto \nu^{-p/2}t^{-p}.
\end{array}
\end{equation}
When the energy of accelerated electrons are nearly in equipartition
with protons, we have
$\epsilon_e\sim 1$, where the jet evolution in the fully-radiative and
spherical-like phase, $\gamma\propto t^{-6/13}$, is closer to that of
the spreading phase, $\gamma\propto t^{-1/2}$. So we can expect that the
light-curve break due to sideways expansion
is less obvious. In fact, if $\epsilon_e=0.6$,
which is the estimated value for GRB 970508 by Granot et al. (1999), and $p=2.4$,
the steepening around the break time from $t^{-1.9}$ to $t^{-2.4}$ is really weak.
Furthermore, the numerical calculations of jetted afterglows by some authors
(e.g., Moderski et al. 1999; Wei \& Lu 2000; Huang et al. 2000a,b)
showed a smooth change in light curves, and thus
we expect that there may be no light curve breaks observed in the
case discussed here.
That is to say, no break dosen't mean a burst without beaming.
Table 2 summarizes the $\alpha-\beta$ relation above $\nu_m$ before and
after the break time $t_{jet}$. Fig. 5 depicts further the relation above
$\nu_c$.

\begin{center}
\begin{table*}
\begin{center}
\begin{tabular}{|c||c||c|c|}
\hline
& spectral index & \multicolumn{2}{|c|}{light curve index $\alpha$
($F_{\nu}\propto t^{-\alpha}$)} \\
& $\beta$ ($F_{\nu}\propto \nu^{-\beta}$) & sphere & jet \\ \hline\hline
&  & $\alpha=(m-2)/(1+m)$ &  \\
\raisebox{1.5ex}[0pt]{$\nu_c<\nu<\nu_m$} & \raisebox{1.5ex}[0pt]{$1/2$}
& $\alpha=1/4$ --- $10/13$ & \raisebox{1.5ex}[0pt]{$\alpha=1$} \\ \hline
&  & $\alpha=(mp-2)/(1+m)$ & $\alpha=p$ \\
\raisebox{1.5ex}[0pt]{$\nu>\nu_m$} & \raisebox{1.5ex}[0pt]{$p/2$}
& $\alpha=(3\beta-1)/2$ --- $(24\beta-2)/13$
& $\alpha=2\beta$ \\ \hline
\end{tabular}
\end{center}
\par
\label{t:afterglow}
\caption{Same as Table 1 but in the case of a radiative jet break.
Note that $m=$ 3---12 has been used to yield the range of $\alpha$. }
\end{table*}
\end{center}

\subsubsection{Slow cooling}
Finally, when $\nu_m=\nu_c$, the electrons become slow cooling, the temporal
scalings of the break frequencies and the fluxes are the same as those in
the later spreading phase of an adiabatic jet in section 3.2.2. So Eq. (28)
are valid here. But the $t_{cm}$ calculated in section 3.2.2 is not valid here
because the jet has been in the spreading phase rather than the spherical-like phase.
This time becomes
\begin{equation}
t_{cm}^r=t_{jet}\left (\frac{\nu_{m,jet}}{\nu_{c,jet}}\right )^{1/2}
\end{equation}
(see a detail in Appendix),
where the subscript ``jet" denotes the quantities in time $t_{jet}$.

\section{Summary and Discussion}
We have analyzed the whole process of a relativistic jet expanding into a homogeneous
environment, from the radiative to adiabatic regime and from the spherical-like
to spreading phase. Then we have found two different kinds of light-curve break
for the jet effect.
One is an adiabatic jet break which appears if the transition of the jet
into the spreading phase happens in the adiabatic regime
when the accelerated electrons are already slow cooling.
This case is widely discussed already.
Another is a radiative jet break which corresponds to the jet spreading obviously
in the radiative regime when the  accelerated electrons are still fast cooling.
This case leads to a weaker break in the light curve if $\epsilon_e\sim 1$.

Based on the dynamics, we have derived the light curves of synchrotron emission
from jetted
afterglows, and summarized our main results as follows. First, in the earliest
radiative and spherical-like phase, the light curve decay index is steep and
relevant to $\epsilon_e$ shown in Eq. (20),
e.g., the flux at high energy $\nu>\nu_m$ decays as fast
as  $t^{-2}$ for $\epsilon_e\simeq 1$ and $p\simeq 2.4$. Thus,
as pointed out by B\"ottcher \& Dermer (2000), the rapid detections of
afterglows at early times will provide a way to determine the electron energy
fraction $\epsilon_e$ of total energy density.

Second, if the Compton cooling of electrons dominates the synchrotron cooling,
the afterglows will exhibit quite different behaviors from the previous
predictions. In the case of an adiabatic jet break, the sharp steepening of light
curve in both $\nu<\nu_c$ and $\nu>\nu_c$
is only expected for the case of $\epsilon_e/\epsilon_B\ll 1$,
i.e. $Y<1$ in the beginning. While for the case of $\epsilon_e/\epsilon_B\gg1$,
the jet begins
with strong Compton scattering ($Y>1$), and it might dominate total cooling
over the whole relativistic stage (Sari \& Esin 2000).
In this case, the light curve steepening in $\nu<\nu_c$ is accompanied
with a flattening (or even a climb) in $\nu>\nu_c$,
due to the increasing cooling frequency $\nu_c$ for $Y>1$.

Finally, in the case of a radiative jet break, the late-time change of
the dynamical evolution due to jet effect is weaker than
in the case of an adiabatic jet break,
which results in a weaker light curve break.
If $\epsilon_e\sim 1$, the change is practically smooth, yielding a
smooth ``break''.
Thus, this kind of jet will show a singly steep light curve
without obvious steepening break in $\nu>\nu_c$,
even though the jet is highly collimated with a small $\theta_0$.

A few GRB afterglows are observed to have an achromatic light
curve break, e.g., GRB 990123 (Kulkarni et al. 1999; Castro-Tirado
et al. 1999; Fruchter et al. 1999), GRB 990510 (Harrison et al.
1999; Stanek et al. 1999), GRB 991216 (Halpern et al. 2000), GRB
000301C (Rhoads \& Fruchter 2001; Masetti et al. 2000; Jensen et
al. 2001; Berger et al. 2000; Sagar et al. 2000), GRB 000418
(Berger et al. 2001), and GRB 000926 (Price et al. 2001; Harrison
et al. 2001; Sagar et al. 2001a; Piro et al. 2001), GRB 010222
(Masetti et al. 2001; Stanek et al. 2001; Sagar et al. 2001b;
Cowsik et al. 2001; In 't Zand et al. 2001). Jets in GRBs are
usually proposed to account for these breaks. However, in view of
our work, two cases of relativistic jets are ruled out to explain
these breaks: (i) an adiabatic jet ($t_{cm}<t_{jet}$) with strong
Compton cooling ($Y>1$) and the cooling frequency $\nu_c$ locating
in the observed energy range; (ii) a radiative jet
($t_{cm}>t_{jet}$) with a significant fraction of the total energy
occupied by electrons ($\epsilon_e \sim 1$).

It should be noted that all the discussions here are concerned
about the relativistic stage. If a relativistic jet expands into a
medium as dense as $10^3-10^6\,\,{\rm cm}^{-3}$, such as a
circumstellar cloud (Galama \& Wijers 2001), the blast wave must
enter the non-relativistic stage within a few days, leading to a
steepening of the afterglow light curve (Dai \& Lu 1999, 2000).
This transition from the relativistic to non-relativistic stage
produces another promising explanation for the broken afterglow
light curves.

\section*{Acknowledgments}
Z. Li would like to thank D. M. Wei, X. Y. Wang and Y. F. Huang
for valuable discussions. The authors thank the anonymous referee
for valuable and detailed comments. This work was supported by the
National Natural Science Foundation of China under grants 19973003
and 19825109, and the National 973 project (NKBRSF G19990754).

\section*{Appendix}
The time-evolution of the three break frequencies and peak flux in the earliest
radiative and spherical-like phase are given below:
\begin{eqnarray}
\nu_m&=&f(k,m)\left (\frac{p-2}{p-1}\right )^2(1+X)^{-2}(1+z)^{1+\frac{k-4}{2(1+m)}}
 x_pA^{\frac{k+m-3}{2(k-3)(1+m)}}E_0^{\frac{(k-4)m}{2(k-3)(1+m)}}\gamma_0^
{-\frac{(k-4)(k+m-3)}{(k-3)(1+m)}} \nonumber \\
& \times & \epsilon_B^{1/2}\epsilon_e^2
t^{-\frac{k+4m}{2(1+m)}}{\rm Hz},
\end{eqnarray}
where
\begin{equation}
f_m(k,m)=2.88\times10^{13}\times 53.32^{\frac{(k-4)m}{2(k-3)(1+m)}}(6.00\times10^{10})^
{-\frac{k+4m}{2(1+m)}}(3-k)^{\frac{(k-4)m}{2(k-3)(1+m)}};
\end{equation}
\begin{equation}
\nu_c=f(k,m)(1+Y)^{-2}(1+z)^{\frac{2-3k-2m}{2(1+m)}}A^{\frac{9-3k+5m}{2(k-3)(1+m)}}
E_0^{\frac{(4-3k)m}{2(k-3)(1+m)}}
\gamma_0^{\frac{(3k-4)(k+m-3)}{(k-3)(1+m)}}\epsilon_B^{-3/2}
t^{-\frac{4-3k}{2(1+m)}}{\rm Hz},
\end{equation}
where
\begin{equation}
f_c(k,m)=3.71\times10^{45}(3-k)^{\frac{(4-3k)m}{2(k-3)(1+m)}}(2+m)^2
10^{\frac{64.67+k(16.17k-2.59m-70.06)+3.45m}{(k-3)(1+m)}};
\end{equation}
\begin{eqnarray}
\nu_a&=&f_a(k,m)(1+X)^{3/5}(1+Y)(1+z)^{\frac{9k+m-13}{5(1+m)}}
A^{\frac{9k-13m-27}{5(k-3)(1+m)}}E_0^{\frac{(9k-14)m}{5(k-3)(1+m)}}
\gamma_0^{-\frac{2(9k-14)(k+m-3)}{5(k-3)(1+m)}} \nonumber \\
& \times &\epsilon_B^{6/5}t^{-\frac{9k+6m-8}{5(1+m)}}{\rm Hz},
\end{eqnarray}
where
\begin{equation}
f_a(k,m)=4.05\times10^{-19}(3-k)^{\frac{(9k-14)m}{5(k-3)(1+m)}}(2+m)^{-1}
10^{\frac{k(75.44-19.40k-9.83m)+33.96m-51.73}{(k-3)(1+m)}};
\end{equation}
\begin{eqnarray}
F_m&=&f_F(k,m)\phi_pA^{\frac{3k-m-9}{2(k-3)(1+m)}}
(1+X)(1+z)^{\frac{3k+4m-4}{2(1+m)}}E_0^{\frac{(3k-8)m}{2(k-3)(1+m)}}
\gamma_0^{-\frac{(3k-8)(k+m-3)}{(k-3)(1+m)}} \epsilon_B^{1/2}\nonumber \\
& \times &d_l^{-2}t^{-\frac{3k+2m-6}{2(1+m)}}
{\rm {erg \,s}}^{-1}{\rm cm}^{-2},
\end{eqnarray}
where
\begin{equation}
f_F(k,m)=2.42\times10^{-21}[53.32(3-k)]^{\frac{6+k(m-2)-2m}{2(k-3)(1+m)}}
10^{\frac{32.33-16.17k-10.78m}{1+m}}.
\end{equation}

The time when $\nu_m=\nu_c$ is given by
\begin{eqnarray}
t_{cm}&=&f_{cm}(k,m)\left [\frac{(1+Y)(p-2)}{(1+X)(p-1)}\right ]^{\frac{1+m}{k+m-1}}
(1+z)x_p^{\frac{1+m}{2(k+m-1)}}A^{\frac{k-m-3}{(k-3)(k+m-1)}} \nonumber \\
 &\times &\ E_0^{\frac{(k-2)m}{(k-3)(k+m-1)}}\gamma_0^{-\frac{2(k-2)(k+m-3)}{(k-3)(k+m-1)}}
 (\epsilon_B\epsilon_e)^{\frac{1+m}{k+m-1}},
\end{eqnarray}
where
\begin{equation}
f_{cm}(k,m)=1.67\times10^{-11}\left (\frac{8.82\times10^{-17}}{2+m}\right )^
{\frac{1+m}{k+m-1}}[53.32(3-k)]^{\frac{(k-2)m}{(k-3)(k+m-1)}}.
\end{equation}
for the case of an adiabatic break $(t_{cm}<t_{jet})$, and by
\begin{equation}
t_{cm}^r=f_{cm}^r(m)\frac{(1+Y)(p-2)}{(1+X)(p-1)}(1+z)x_p^{1/2}A^{1/3}E_0^{2/3}
\gamma_0^{-{4\over3}+{4\over m}}\epsilon_B\epsilon_e\theta_0^{4/m}{\rm s},
\end{equation}
where
\begin{equation}
f_{cm}^r(m)=1.47\times10^{-27}\frac{(4.91\times10^{-10})^{\frac1{1+m}}}{2+m}
10^{\frac{10.78+1.47m}{1+m}}.
\end{equation}
for the case of a radiative break $(t_{jet}^a<t_{cm})$. The break time when a relativistic
jet begins to spread exponentially is given by
\begin{equation}
t_{jet}=f_{jet}(k,m)(1+z)A^{1\over k-3}E_0^{1\over k-3}\gamma_0^{2(\frac 1{k-3}+\frac 1m)}
\theta_0^{2+\frac 2m}{\rm s},
\end{equation}
where
\begin{equation}
f_{jet}(k,m)=(5.22\times10^{-24})^{1\over k-3}(6.00\times10^{10})^{5-k\over k-3}(3-k)
^{1\over 3-k}.
\end{equation}
for the case of a radiative break, and by
\begin{eqnarray}
t_{jet}^a&=&f_{jet}^a(k,m)
\left [\frac{(1+Y)(p-2)}{(1+X)(p-1)}\right ]^{-{m-3\over3(m-1)}}
(1+z)^{3-m\over3(1+m)}x_p^{-\frac{m-3}{6(m-1)}} \nonumber \\
& \times &A^{-{5m-9\over 9(m-1)}}E_0^{2m\over 9(m-1)}
\gamma_0^{-\frac {4(m-3)}{9(m-1)}}(\epsilon_B\epsilon_e)^{-{m-3\over3(m-1)}}
\theta_0^{8/3}{\rm s},
\end{eqnarray}
where
\begin{equation}
f_{jet}^a(k,m)=4.17\times10^{-12}159.96^{2m\over9(m-1)}
\left (\frac{8.82\times10^{-17}}{2+m}\right )^{-{m-3\over3(m-1)}}(1+m).
\end{equation}
for the case of an adiabatic break.

\newpage
\begin{figure}
\centerline{\hbox{\psfig{figure=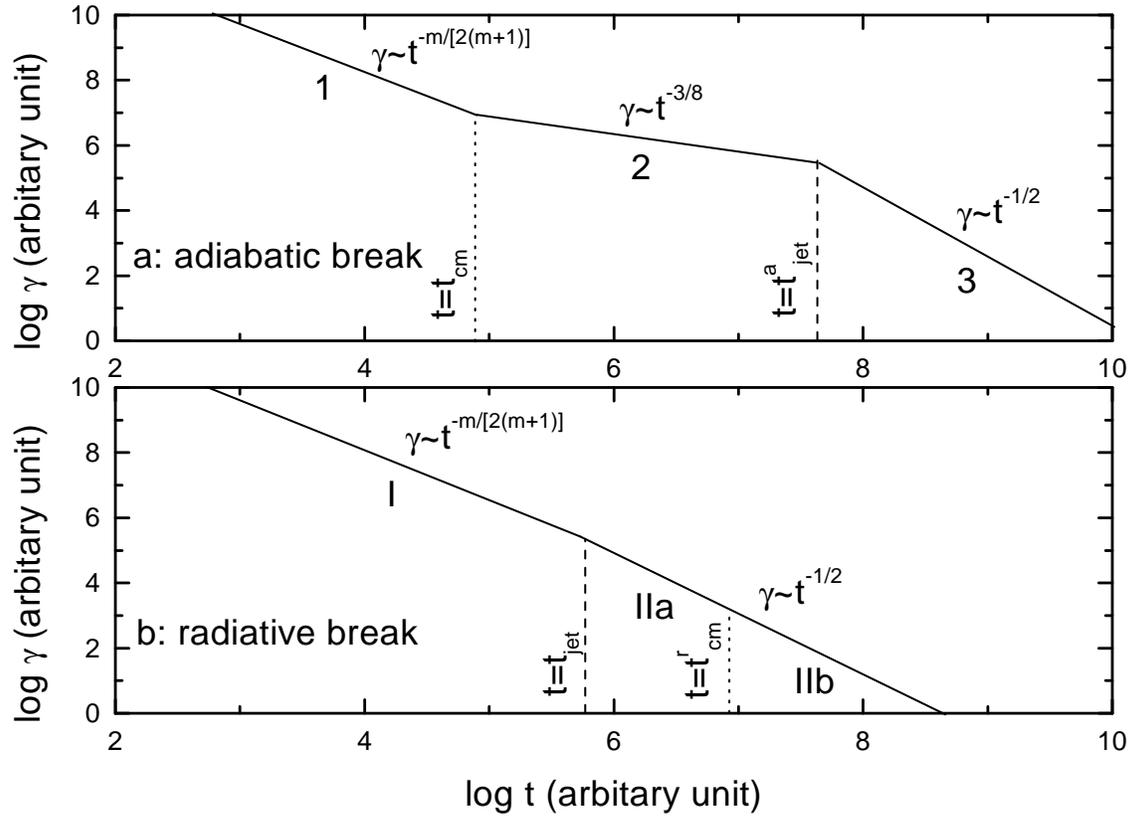,width=7.0in,angle=0}}}
\caption {The Lorentz factor of a relativistic jet as a function of the observer's
time (schematic). In the case of an adiabatic jet break ({\em frame {\bf a}}),
the relativistic jet evolves sequentially into
(1) a radiative and spherical-like phase, (2) an adiabatic and spherical-like phase and
(3) a spreading phase. In the case of a radiative jet break ({\em frame {\bf b}}),
the sequence becomes (I) a radiative and spherical-like phase and (II) a spreading phase
which consists of two stages with fast-cooling (II{\it a})
and slow-cooling (II{\it b}) electrons, respectively.
The times separating different phases are illustrated in the text.}
\end{figure}

\newpage
\begin{figure}
\centerline{\hbox{\psfig{figure=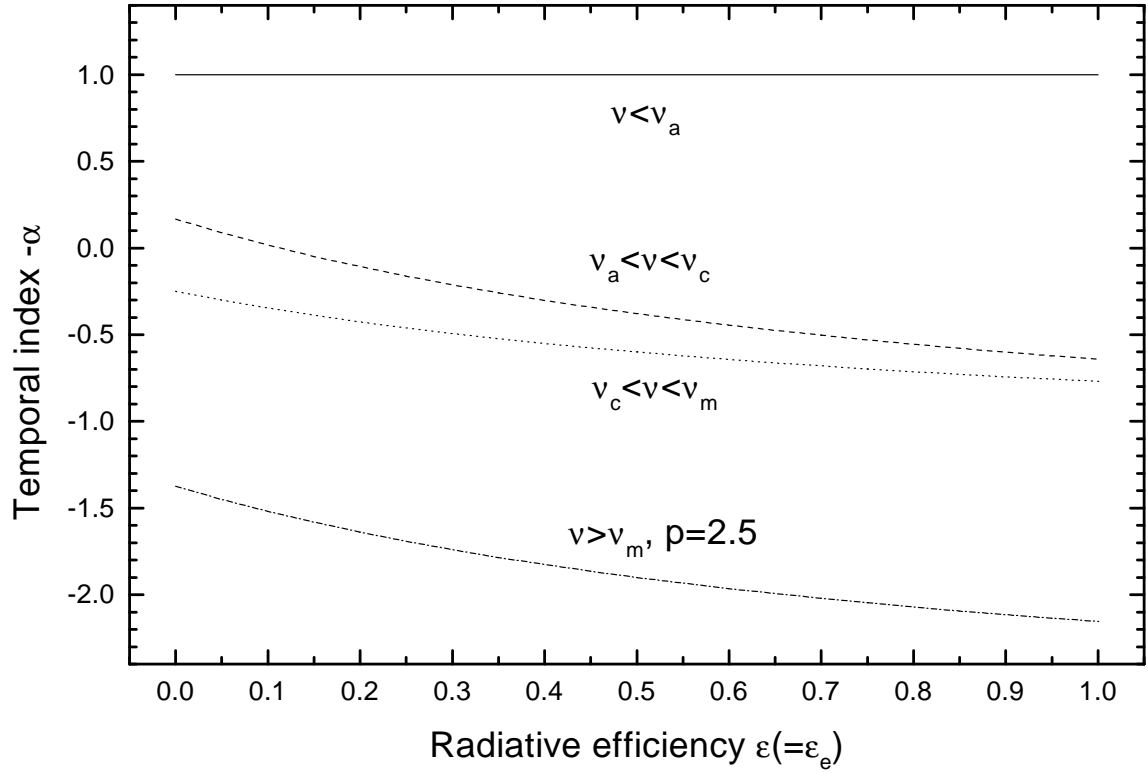,width=7.0in,angle=0}}}
\caption {The relation between the light curve decay index $-\alpha$
and the radiative efficiency $\epsilon(=\epsilon_e)$ in different energy bands
for the spherical-like and radiative phase.}
\end{figure}

\newpage
\begin{figure}
\centerline{\hbox{\psfig{figure=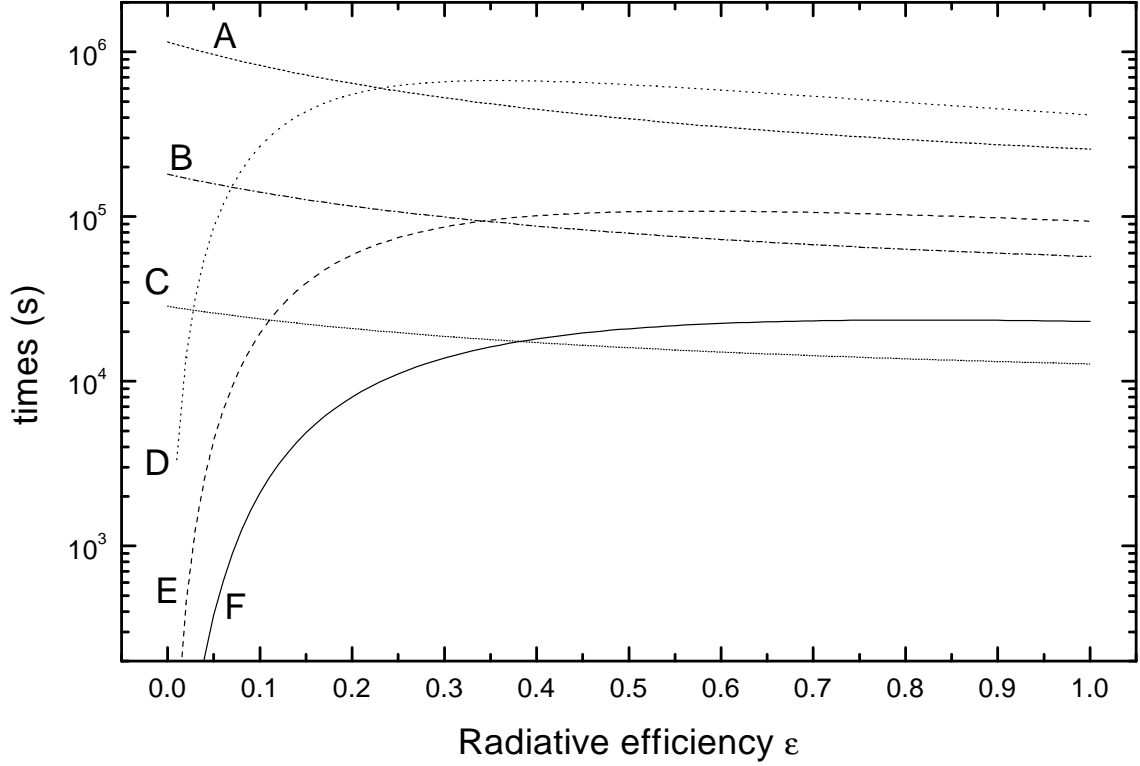,width=7.0in,angle=0}}}
\caption {The times $t_{cm}$ and $t_{jet}$ as functions of radiative efficiency
$\epsilon(=\epsilon_e)$ with different parameters: curves A, B and C show the
time $t_{jet}$ with $\theta_0=0.2$, 0.1 and 0.05, respectively; curves D, E and F
show the time $t_{cm}$ with $\epsilon_B=10^{-1}$, $10^{-2}$ and $10^{-3}$,
respectively. The remaining parameters are set as $k=0$, $X=1$, $z=1$,
$A=1\,{\rm cm}^{-3}$, $p=2.5$, $E_0=10^{53}$erg, and $\gamma_0=100$.}
\end{figure}

\newpage
\begin{figure}
\centerline{\hbox{\psfig{figure=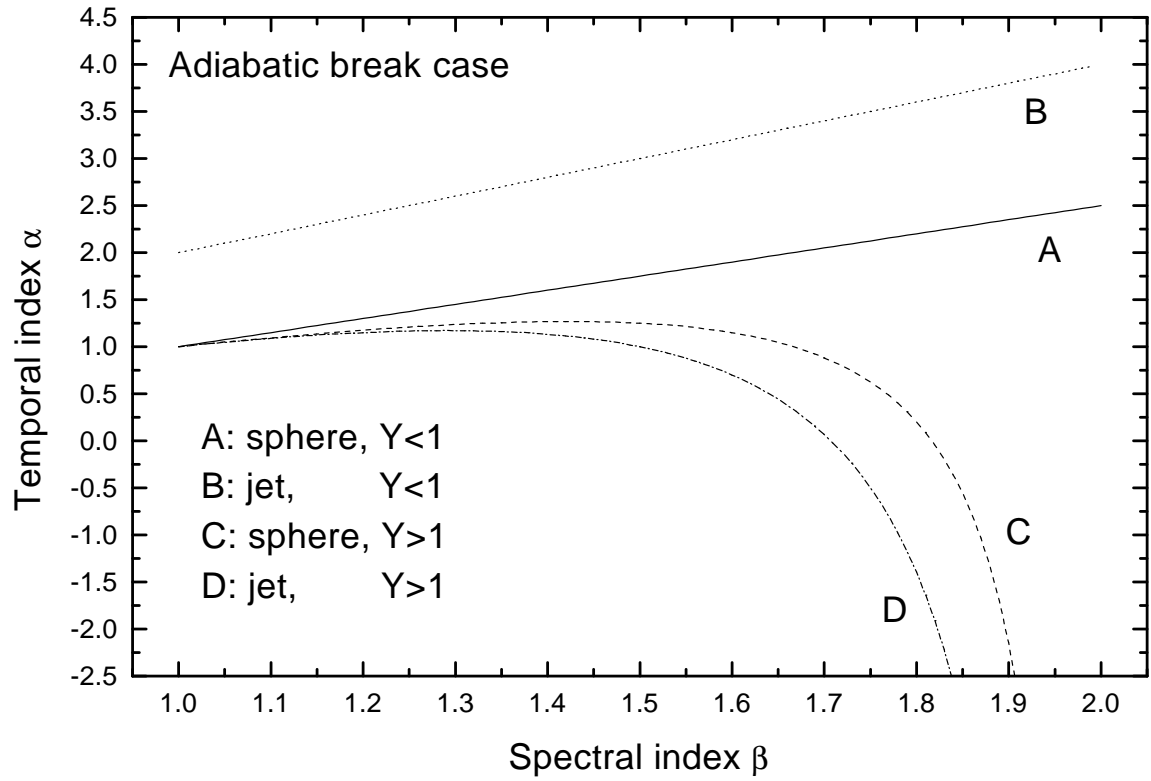,width=7.0in,angle=0}}}
\caption {The relation between the light curve index $\alpha$ and the
spectral index $\beta$ above $\nu_c$ in the case of an adiabatic jet break.
The cases A and C
correspond to times before the jet break, while B and D correspond
to times after the jet break.}
\end{figure}

\newpage
\begin{figure}
\centerline{\hbox{\psfig{figure=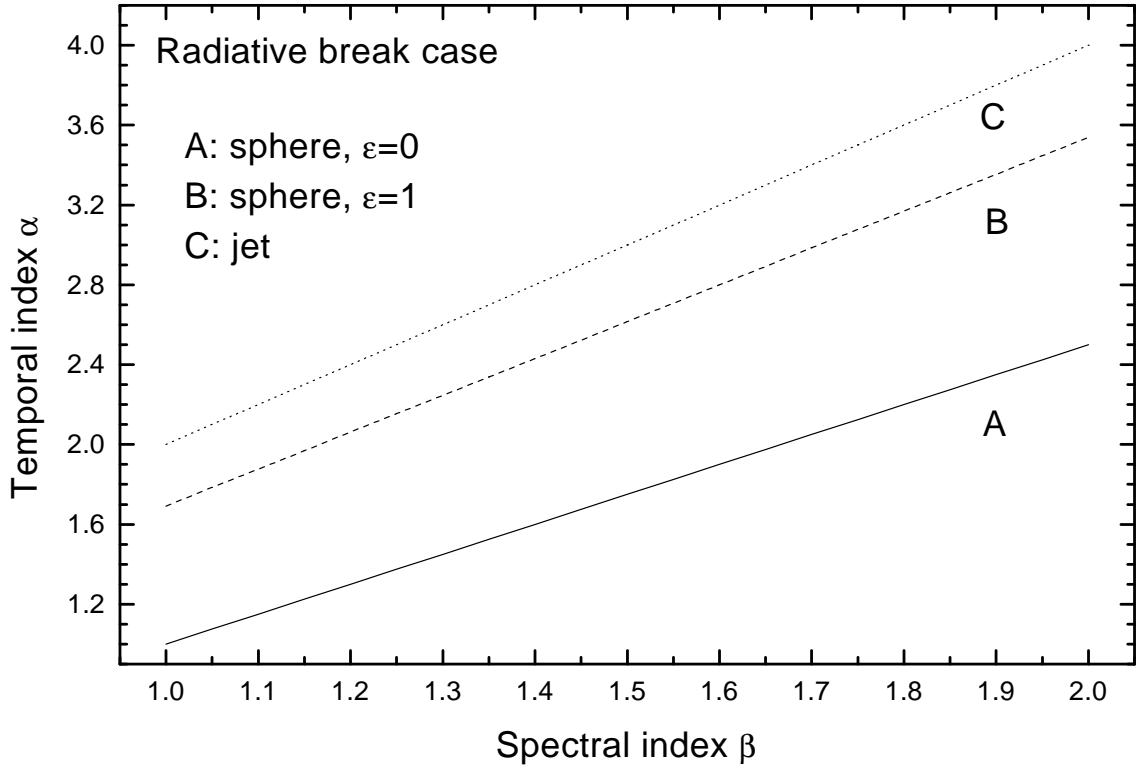,width=7.0in,angle=0}}}
\caption {Same as Fig. 4 but at the energy band above $\nu_m$ in
the case of a radiative jet break. Curves A and B correspond to the spherical-like
phase with $\epsilon=0$ and 1, respectively. The cases for $0<\epsilon<1$ are
located in the regime between A and B.
Curve C corresponds to the spreading phase.}
\end{figure}

\end{document}